\documentclass[aps,prd,twocolumn,floatfix,10pt,amssymb,amsfont,amsmath,nofootinbib,float,floatfix,longbibliography]{revtex4-1}

\usepackage[utf8]{inputenc}
\usepackage{lmodern}
\usepackage[american]{babel}
\usepackage[autostyle=try,english=american]{csquotes}
\usepackage{xfrac}
\usepackage{microtype}
\usepackage{epstopdf}
\usepackage{verbatim}
\usepackage[usenames,dvipsnames]{xcolor}
\usepackage{graphicx}
\usepackage[pdftex,plainpages=false]{hyperref}
\usepackage{todonotes}
\usepackage{mathtools}
\usepackage{array}
\usepackage{tabularx}
\usepackage{booktabs}
\usepackage[export]{adjustbox}
\usepackage{bbm}
\usepackage{enumitem}
\usepackage{gensymb}
\usepackage{amsthm}

\hypersetup{%
	bookmarks=true,         
	bookmarksopen=true,
	unicode=true,           
	pdftoolbar=true,        
	pdfmenubar=true,        
	pdffitwindow=false,     
	pdfstartview={FitH},    
	pdftitle={Uhlmann fidelities from tensor networks},
	pdfauthor={M. Hauru, G. Vidal},
	pdfsubject={},          
	pdfcreator={},          
	pdfproducer={pdfLaTeX}, 
    pdfkeywords={matrix product states} {tensor networks} {fidelity},
	pdfnewwindow=true,      
	colorlinks,
	citecolor=Maroon,
	filecolor=Maroon,
	linkcolor=Maroon,
	urlcolor=Maroon
}

\DeclareMathOperator{\Tr}{Tr}

\DeclareMathOperator{\rank}{rank}

\newcommand{\bra}[1]{\langle #1|}
\newcommand{\ket}[1]{|#1\rangle}
\newcommand{\braket}[2]{\langle #1|#2\rangle}

\newcommand{\unity}{\mathbbm{1}}

\newcommand{\dg}{^{\dagger}}

\newtheorem{thrm:generic_purification}{Theorem}
\newcommand{\Fsep}{F_d}
\newcommand{\statespace}{\mathcal{H}}
\newcommand{\ancillaspace}{\mathcal{H}_{\text{anc.}}}

\graphicspath{{./fig/}}

\begin{document}

\title{Uhlmann fidelities from tensor networks}

\author{Markus Hauru}
\email{markus@mhauru.org}

\author{Guifre Vidal}

\affiliation{Perimeter Institute for Theoretical Physics, Waterloo, Ontario N2L 2Y5, Canada}
\affiliation{Department of Physics and Astronomy, University of Waterloo, Waterloo, Ontario N2L 3G1, Canada}

\date{\today}

\begin{abstract}
    Given two states $\ket{\psi}$ and $\ket{\phi}$ of a quantum many-body system, one may use the overlap or fidelity $|\braket{\psi}{\phi}|$ to quantify how similar they are.
    To further resolve the similarity of $\ket{\psi}$ and $\ket{\phi}$ in space, one can consider their reduced density matrices $\rho$ and $\sigma$ on various regions of the system, and compute the Uhlmann fidelity $F(\rho, \sigma) = \Tr \sqrt{\sqrt{\rho} \sigma \sqrt{\rho}}$.
    In this paper, we show how computing such subsystem fidelities can be done efficiently in many cases when the two states are represented as tensor networks.
    Formulated using Uhlmann's theorem, such subsystem fidelities appear as natural quantities to extract for certain subsystems for Matrix Product States and Tree Tensor Networks, and evaluating them is algorithmically simple and computationally affordable.
    We demonstrate the usefulness of evaluating subsystem fidelities with three example applications: studying local quenches, comparing critical and non-critical states, and quantifying convergence in tensor network simulations.
\end{abstract}

\maketitle


\section{Introduction}%
\label{sec:introduction}
Given two, pure many-body quantum states $\ket{\psi}$ and $\ket{\phi}$, their similarity can be quantified by their fidelity $|\braket{\psi}{\phi}|$.
It is intuitively clear, however, that there is more to say:
One can discuss the similarity of the two states with regard to certain parts of the system, and make statements such as ``the two states are similar at short length scales, but not at long length scales'', or ``the two states only significantly differ from each other in this particular region''.
This intuition is quantified by the Uhlmann fidelity of the reduced states of $\ket{\psi}$ and $\ket{\phi}$ on the subsystems in question, which for two density matrices $\rho$ and $\sigma$ is defined as~\cite{uhlmann_1976,nielsen_chuang_2010}
\begin{equation}
    F(\rho, \sigma) = \Tr \sqrt{\sqrt{\rho} \sigma \sqrt{\rho}}.
\end{equation}
When $\rho$ and $\sigma$ are reduced density matrices arising from pure states restricted to a subsystem, we call such fidelities subsystem fidelities.

\begin{figure}[tbp]
    \includegraphics[width=\linewidth]{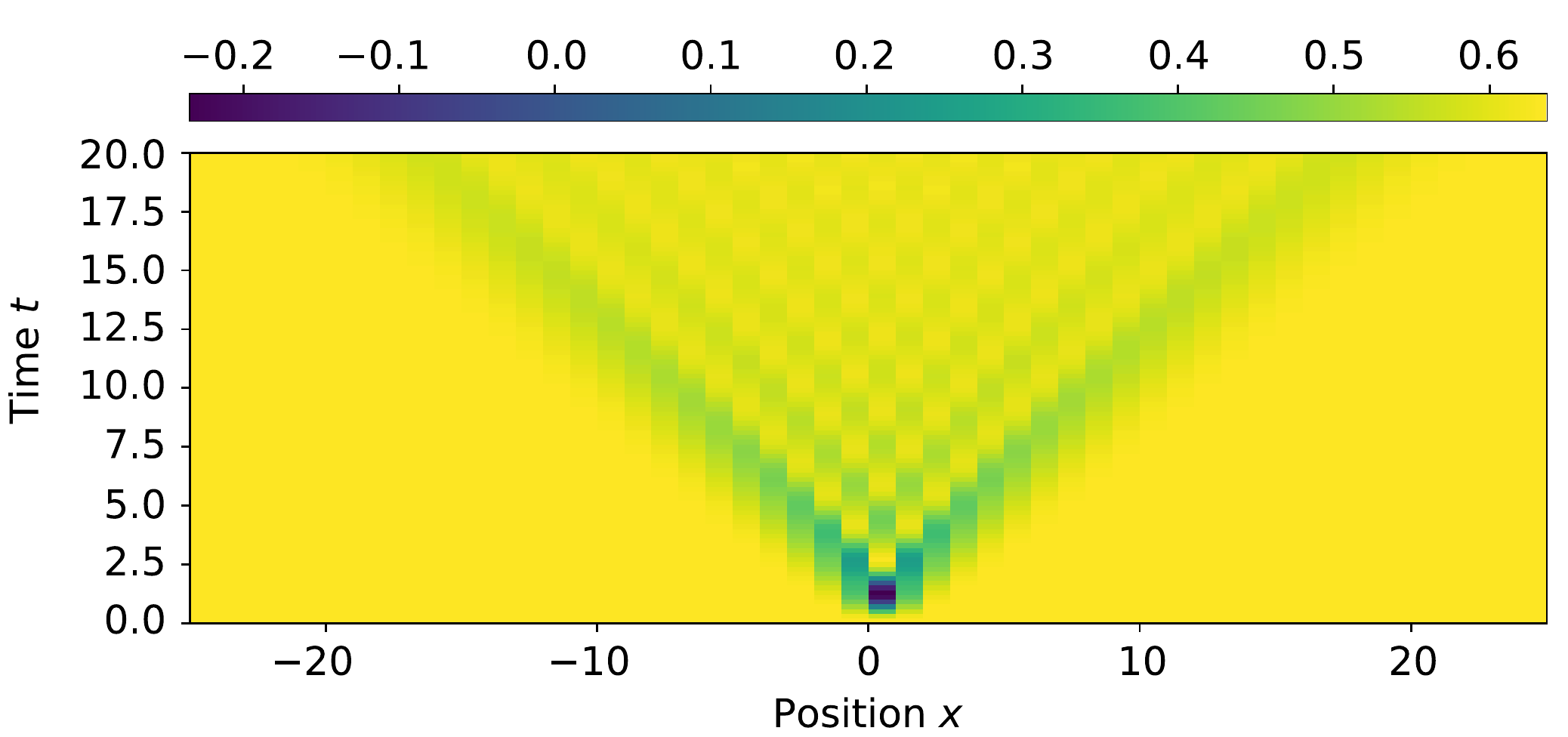}
    \includegraphics[width=\linewidth]{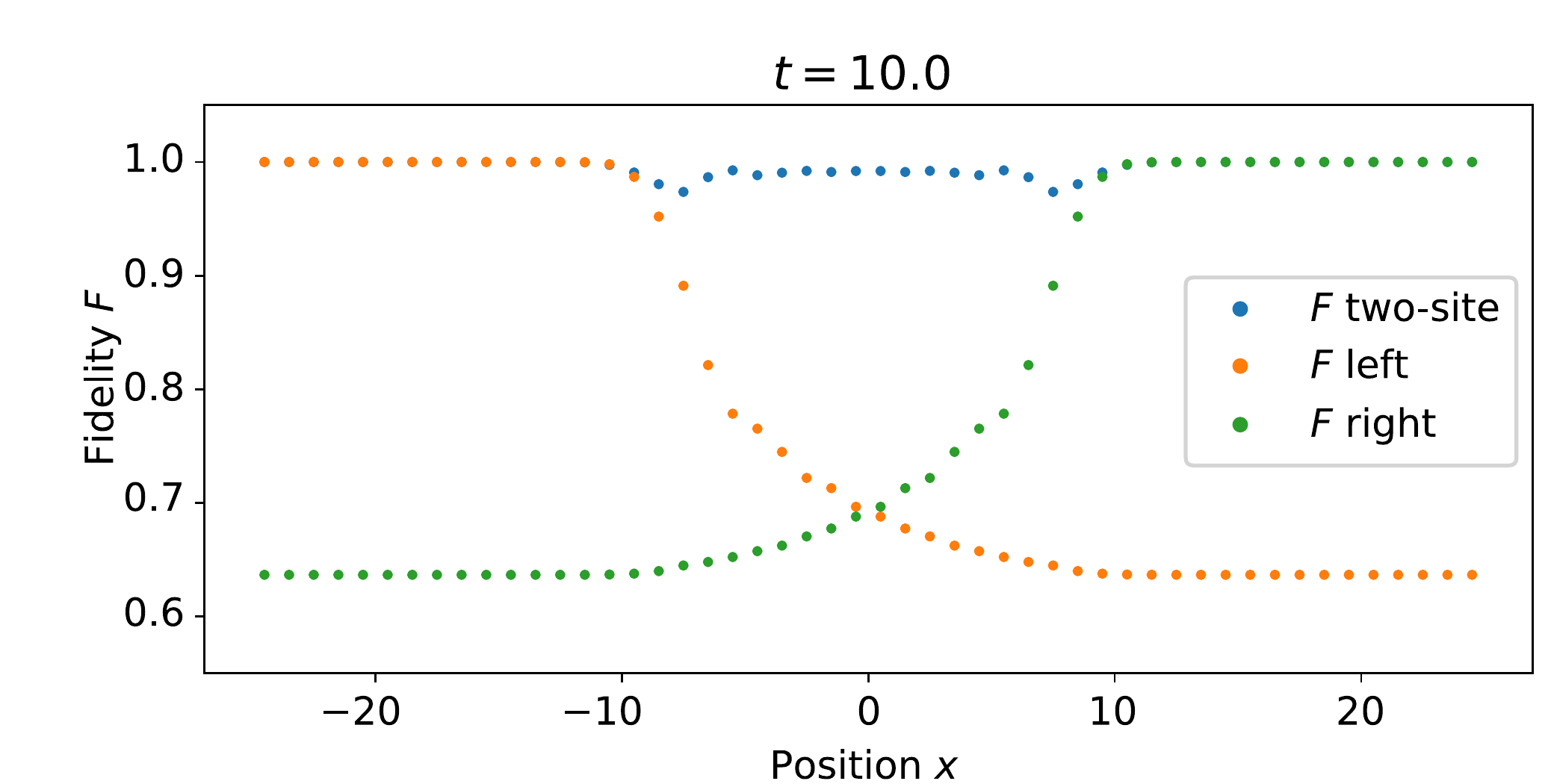}
    \caption{%
        Above, the expectation value of the Pauli $Z$ operator, on the locally quenched state $\ket{\psi(t)}$, as a function of position $x$ and time $t$.
        The model in question is the critical 1D Ising model, and the quench consisted of perturbing the ground state with $Z$.
        This plot serves to merely illustrate the progression of the quench.
        Below, subsystem fidelities between the ground state $\ket{E_0}$ and the quenched state $\ket{\psi(t=10)}$, as functions of position $x$.
        The three different fidelities plotted are the half-system Uhlmann fidelities to the left (descending orange dots) and to the right (ascending green dots) of $x$, and the two-site fidelity at $x$ (blue dots).
        }
    \label{fig:quench_apetizer}
\end{figure}

As an example of a situation where subsystem fidelities are useful quantities to evaluate, consider the following experiment:
Take the ground state $\ket{E_0}$ of a many-body system, disturb it locally with an operator $O$, that could for example flip a single spin, and let the state evolve for some time $t$.
The result of this local quench is an evolved state $\ket{\psi(t)} = e^{itH} O \ket{E_0}$.
In the top half of Fig~\ref{fig:quench_apetizer}, we can see the progression of such a quench in the Ising model, as measured by the magnetization.
The effect of the disturbance can be seen propagating out ballistically.
One might wonder, how the time evolved state $\ket{\psi(t)}$ is different from the ground state $\ket{E_0}$.
It is natural to guess that far away from the local disturbance $\ket{\psi(t)}$ still looks like the ground state, but closer by, the effect of the disturbance has evolved and spread.
This question can be answered by resolving the overlap between $\ket{\psi(t)}$ and $\ket{E_0}$ in space, using subsystem fidelities.
As an example of what such a resolution may look like, in the bottom half of Fig.~\ref{fig:quench_apetizer} we show three different types of subsystem fidelities between $\ket{\psi(t)}$ and $\ket{E_0}$ for the case of the critical 1D Ising model:
One is the fidelity between two-site reduced density matrices, positioned at different places, showing local differences.
The two others are fidelities between reduced density matrices on either the left or right half of the system, where the left-right bipartition is with respect to different points on the lattice.
From the profiles of these fidelities across the spin chain one can clearly see the spread of the disturbance, with the difference between the ground state $\ket{E_0}$ and the quenched state $\ket{\psi(t)}$ being the most prominent at the ballistic front, where the effect of the disturbance is propagating outwards.

Note that to be able to do this comparison between the quenched state and the ground state, we did not need to specify any observables to use as probes, nor any other further information about the system.
Like other quantum information concepts that are nowadays used to analyze many-body systems, such as entanglement entropies, fidelities are entirely agnostic about the nature of the physics in the system, or even the degrees of freedom in question.
Moreover, they are a more sensitive probe than any single observable, in the sense that for any observable to differ between two states, their reduced density matrices on the support of the observable must be different.

To be able to compute subsystem fidelities for many-body states, two main obstacles need to be overcome:
We need a way to efficiently represent many-body states in an exponentially large state space, and second, given representations of two pure states, we need to be able to compute their fidelity on a subsystem of interest.
In this paper we will use tensor network states to overcome these obstacles.
In Sect.~\ref{sec:fidelity_basics} we discuss Uhlmann fidelities in detail, and in particular how they can be formulated in terms of purifications of the reduced density matrices, while avoiding constructing the reduced density matrices themselves.
Then, in Sect.~\ref{sec:evaluating_fidelities}, we turn our attention to tensor network states, which provide an efficient way to describe low-entanglement states of many-body systems, and for many choices of subsystems, also purifications of their reduced density matrices.
We concentrate on Matrix Product States and Tree Tensor Networks, and show in detail how fidelities between two such tensor network states can be evaluated for certain choices of subsystems, at the same leading order computational cost  as producing the states.
Some of the Matrix Product State results in Sect.~\ref{sec:fidelities_mps} were already presented in the appendix of Ref.~\onlinecite{liu_local_2017}, see a note at the end of this paper for more details.
In Sect.~\ref{sec:applications} we return to the above example of a local quench to discuss it in more detail, and present two other applications of subsystem fidelities:
resolving the difference between a critical and an off-critical state as a function of scale, and quantifying convergence and the effects of limited bond dimension in tensor network simulations.
Finally, we conclude in Sect.~\ref{sec:discussion}.

Python 3 source code that implements the Matrix Product State algorithms for evaluating subsystem fidelities described in Sect.~\ref{sec:fidelities_mps} and produces the results shown in Sect.~\ref{sec:applications}, is available at \href{https://arxiv.org/src/1807.01640}{arxiv.org/src/1807.01640}.


\section{Uhlmann fidelity of subsystems}%
\label{sec:fidelity_basics}
Let $\ket{\psi}$ and $\ket{\phi}$ be two states of the same lattice system.
Consider some part of this lattice, call it $M$, and its complement $M^\complement$, and suppose we want to compare $\ket{\psi}$ and $\ket{\phi}$ on $M$.
For this purpose, the natural objects to consider are the reduced density matrices $\rho = \Tr_{M^\complement} \ket{\psi}\bra{\psi}$ and $\sigma = \Tr_{M^\complement} \ket{\phi}\bra{\phi}$, and their similarity can be quantified by their Uhlmann fidelity
\begin{equation}
    \label{eq:fidelity_definition}
    F(\rho, \sigma) = \Tr \sqrt{\sqrt{\rho} \sigma \sqrt{\rho}}.
\end{equation}
The Uhlmann fidelity~\eqref{eq:fidelity_definition} is usually considered to be the most natural generalization of the overlap $|\braket{\psi}{\phi}|$ of pure states to mixed states~\cite{nielsen_chuang_2010}.
It fulfills Jozsa's axioms for fidelities~\cite{jozsa_fidelity_1994}, meaning that it
\begin{itemize}
    \item is symmetric between $\rho$ and $\sigma$
    \item ranges from 0 to 1, and is 1 if and only if $\rho = \sigma$,
    \item is invariant under unitary transformations of the state space of $M$,
    \item reduces to $|\bra{\phi}\rho\ket{\phi}|$ if $\sigma = \ket{\phi}\bra{\phi}$ is pure.
\end{itemize}

Instead of trying to evaluate Eq.~\eqref{eq:fidelity_definition} directly, we will make use of Uhlmann's theorem~\cite{nielsen_chuang_2010}.
It states that for any reduced density matrices $\rho$ and $\sigma$, the Uhlmann fidelity can equivalently be defined as
\begin{equation}
    \label{eq:uhlmann_definition}
    F(\rho, \sigma) = \max_{\ket{\varphi_{\rho}}, \ket{\varphi_{\sigma}}} |\braket{\varphi_\rho}{\varphi_\sigma}|,
\end{equation}
where $\ket{\varphi_\rho}$ and $\ket{\varphi_\sigma}$ are purifications of $\rho$ and $\sigma$, and the maximum is taken over all possible purifications.
As an aside, note at this point, that Uhlmann's theorem makes it obvious that when $\rho$ and $\sigma$ are reduced density matrices arising from pure states restricted to a subsystem $M$, then the Uhlmann fidelity is monotonic in $M$, in that if one increases $M$ to include more of the system, the fidelity must decrease.

To use Eq.~$\eqref{eq:uhlmann_definition}$ to evaluate Uhlmann fidelities, we need to construct the generic form of purifications of $\rho$ and $\sigma$.
A priori this may seem like a daunting task.
However, concentrating for the moment on $\rho$, assume that we have access to a decomposition of the form $\rho = XX\dg$, with some matrix $X$.
This may for instance arise from being able to compute the Schmidt decomposition of $\ket{\psi}$ between $M$ and $M^\complement$, or from some other structure of the state we have access to.
Then, as we review in App.~\ref{app:generic_purification}, all purifications $\varphi_\rho$ of $\rho$, when viewed as matrices between $M$ and the ancilla\footnote{%
    Throughout the paper we often consider bipartite states $\ket{\varphi} \in \statespace{}_1 \otimes \statespace{}_2$ as matrices $\varphi: \statespace{}_1 \to \statespace{}_2$.
    Switching between the two is simply a question of changing between the space $\statespace{}_2$ and its dual, or in other words, the defining relation between $\ket{\varphi}$ and $\varphi$ is that $\bra{i} \varphi \ket{j} = [\bra{i} \otimes \bra{j}] \ket{\varphi}$ for all $\ket{i} \in \statespace{}_1$ and $\ket{j} \in \statespace{}_2$.
    }, can be written in the form
\begin{align}
    \label{eq:generic_purification}
    \varphi_\rho = XW_\rho,
\end{align}
with $W_\rho$ being some isometric matrix, meaning it fulfills $W_\rho W_\rho\dg = \unity$.
Given this, if $\rho = XX\dg$ and $\sigma = YY\dg$, we can write the Uhlmann fidelity between them as
\begin{align}
    F(\rho, \sigma) &= \max_{\ket{\varphi_{\rho}}, \ket{\varphi_{\sigma}}} \left|\braket{\varphi_\rho}{\varphi_\sigma}\right| = \max_{\varphi_{\rho}, \varphi_{\sigma}} \left|\Tr [ \varphi_\rho \varphi_\sigma\dg ] \right|\\
    &= \max_{W_\rho, W_\sigma} \left| \Tr \left[ XW_\rho W_\sigma\dg Y\dg \right] \right|,
    \label{eq:uhlmann_definition_2}
\end{align}
where the last maximum is over all isometries $W_\rho$ and $W_\sigma$.
As we show in App.~\ref{app:one_isometry}, Eq.~\eqref{eq:uhlmann_definition_2} can be further simplified to
\begin{align}
    F(\rho, \sigma) = \max_{W} \left| \Tr \left[ X W Y\dg \right] \right|,
    \label{eq:uhlmann_definition_3}
\end{align}
where $W$ is again an isometry.
Note that the dimensions of $W$ are determined by the dimensions of $X$ and $Y$, and whether $W$'s isometricity means $W W\dg = \unity$ or $W\dg W = \unity$, depends on these dimensions.
Furthermore, the solution to the maximization problem of Eq.~\eqref{eq:uhlmann_definition_3} is straight-forward (see App.~\ref{app:one_isometry}), and given by $W = V U\dg$, where $Y\dg X = U S V\dg$ is the singular value decomposition (SVD) of $Y\dg X$.
This yields for the Uhlmann fidelity, using the cyclicity of trace,
\begin{align}
    F(\rho, \sigma) &= \max_{W} \left| \Tr \left[W Y\dg X \right] \right| = \left| \Tr \left[ V U\dg U S V\dg \right] \right|\\
    &= \left| \Tr S \right| = \lVert Y\dg X \Vert_{\text{tr}},
    \label{eq:uhlmann_definition_4}
\end{align}
with $\lVert \cdot \rVert_{\text{tr}}$ being the trace norm.

Thus we conclude that to be able to compute the fidelity between $\rho$ and $\sigma$, all we need is decompositions of the form $\rho = XX\dg$ and $\sigma = YY\dg$, in such a format that calculating the trace norm of $Y\dg X$ is computationally feasible.
As we shall see in the next section, when the states $\ket{\psi}$ and $\ket{\phi}$ are described as tensor networks, this is often possible.


\section{Evaluating subsystem fidelities from tensor network states}%
\label{sec:evaluating_fidelities}
Tensor networks are classes of many-body states with restricted entanglement structures, that can be efficiently numerically simulated~\cite{Orus:2013kga,Bridgeman:2016dhh}.
They are most easily defined using diagrams such as
\begin{equation}
    \label{eq:MPS_plain}
    \includegraphics[scale=1,raise=-1.1em]{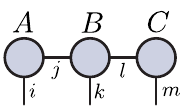} \; = \; \sum_{j,l} A_{ij} B_{jkl} C_{lm}.
\end{equation}
In these diagrams, each node is a tensor, and each link, also known as a ``leg'' or a bond, is an index of that tensor.
Bonds connecting two tensors are contracted over, and free bonds that only have one end point represent physical sites of the system, with values of the free index labeling basis states of the local state space.
Each tensor network diagram defines a class of states in the many-body state space, spanned by all the different choices for the elements of the tensors at each node.
The contracted bonds are typically constrained to range over a finite number of values, called the bond dimension $\chi$, which restricts the states that can be represented by the tensor network.
Typically, the connectivity of the network mirrors the entanglement structure of the states, and each bond can be roughly speaking seen as having the capacity to carry $\log \chi$ bits of entanglement.
The restricted connectivity of the network, and limits on the bond dimension guarantee that tensor network states can be efficiently manipulated numerically, optimizing the elements of each tensor to represent a desired state, and extracting observables from the state.

Nowadays, tensor network methods are the dominant numerical method for studying 1D quantum lattice models~\cite{white_realtime_2004,verstraete_mpsreview_2008,schollwoeck_densitymatrix_2011,0295-5075-24-4-010,PhysRevLett.93.227205,daley_tebd_2004,doi:10.1063/1.1449459,vidal_class_2008,shi_classical_2006} and a strong candidate for the state-of-the-art for many models in two dimensions~\cite{2011arXiv1105.1374S,Verstraete:2004cf,2012PhRvX2d1013C,2014PhRvL.113d6402C}.
The most prominent class of tensor network states is that of Matrix Product States (MPS)~\cite{white_density_1992,PhysRevB.48.10345,rommer_dmrgmps_1997,Fannes1992,dukelsky_dmrgmps_1997,mps_representations_2007}, that have a linear structure like the one above in Eq.~\eqref{eq:MPS_plain}.
MPSes are well suited for describing states of 1D systems that obey the area law of entanglement, most notably low energy states of gapped Hamiltonians.
Other notable classes are the higher dimensional generalization of MPS, called Tensor Product States (TPS) or Project Entangled Pair States (PEPS)~\cite{nishio_tps_2004,Verstraete:2004cf}, as well as Tree Tensor Networks (TTN)~\cite{shi_classical_2006}, and the Multiscale Entanglement Renormalization Ansatz (MERA)~\cite{PhysRevLett.101.110501}, which are both based on notions of coarse-graining, and used mainly for 1D systems.
All tensor network methods share the advantages that they are fully non-perturbative and do not suffer from the sign problem of Monte Carlo methods, making them equally applicable to systems of strong and weak interactions, and bosons and fermions~\cite{corboz_fermionic_2009}.

In this section of the paper, we describe how many subsystem fidelities can be easily evaluated for states that are described as Matrix Product States or Tree Tensor Networks, using Uhlmann's theorem as explained in Sect.~\ref{sec:fidelity_basics}.

\subsection{Matrix Product States}%
\label{sec:fidelities_mps}
The most widely used type of tensor network is the Matrix Product State (MPS), which represents states of 1D lattice systems that respect the area law.
Here we show how to evaluate subsystem fidelities between two MPSes, for two different choices of the subsystem: the left (or right) side of a system partitioned at some point $x$, and a finite window between two points $x_0$ and $x_1$.

\subsubsection{Left-right bipartitions}%
\label{sec:halfsystem_fidelities_for_mps}
Let $\ket{\psi}$ and $\ket{\phi}$ be two Matrix Product States, given by the MPS tensors $A^{(n)}$ and $B^{(n)}$, where $n$ labels lattice sites.
Let $x$ be a point of the lattice, that divides it into a left ($L$) and a right ($R$) half (not necessarily of the same size).
Using $\ket{\psi}$ as an example, this can be expressed graphically as
\begin{equation}
    \label{eq:MPS_A}
    \ket{\psi} = \includegraphics[scale=1,raise=-2.3em]{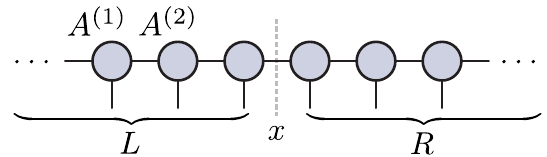} \;.
\end{equation}
We now ask what is the Uhlmann fidelity between the reduced density matrices $\rho_L = \Tr_R \ket{\psi}\bra{\psi}$ and $\sigma_L = \Tr_R \ket{\phi}\bra{\phi}$, that describe the left half of the system.
Here the MPSes may be finite with open boundaries, semi-infinite, or infinite.
We concentrate on the left half and call $\rho_L = \rho$ and $\sigma_L = \sigma$, but the right half can be treated the same way.

Let us concentrate on finding a decomposition $\rho = XX\dg$, as discussed in Sect.~\ref{sec:fidelity_basics}.
Given an MPS like the one in Eq.~\eqref{eq:MPS_A}, one can follow a standard procedure~\cite{mps_representations_2007,itebd_2008} to gauge transform it, i.e.\ to insert partitions of the identity on the contracted indices, to put it into the canonical form
\begin{equation}
    \label{eq:MPS_A_canonical}
    \ket{\psi} = \includegraphics[scale=1,raise=-1.0em]{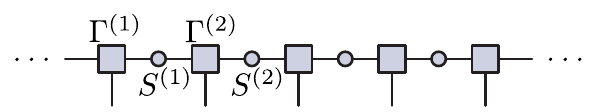} \;.
\end{equation}
Here $S^{(n)}$'s are diagonal matrices with the Schmidt values of the left-right bipartition at $n$ on the diagonal, and together with the $\Gamma^{(n)}$'s they fulfill the orthogonality conditions
\begin{equation}
    \label{eq:MPS_orthogonality}
    \includegraphics[scale=1,raise=-1.0em]{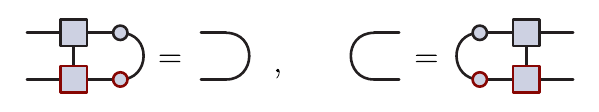} \;.
\end{equation}
Here and in many equations later on, red boundaries denote complex conjugation of the tensor.
The orthogonality condition~\eqref{eq:MPS_orthogonality} guarantees that
\begin{align}
    \rho \; &= \; \includegraphics[scale=1,raise=-2.05em]{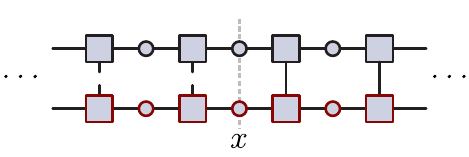}\\
    &= \; \includegraphics[scale=1,raise=-2.05em]{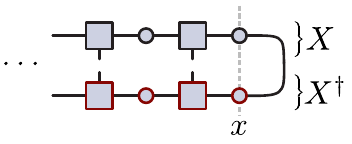} \;.
\end{align}
This is of the desired form $\rho = XX\dg$ as indicated above, and thus invoking Eq~\eqref{eq:generic_purification}, we know that every purification of $\rho$ can be written as
\begin{equation}
    \label{eq:MPS_A_canonical_w_v2}
    \includegraphics[scale=1,raise=-1.0em]{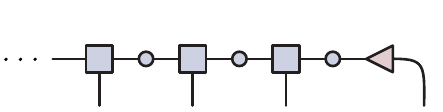} \;,
\end{equation}
where \includegraphics[scale=1, valign=c, raise=-0.05em]{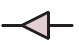} $ = W_\rho$ from Eq.~\eqref{eq:generic_purification}, and the right-most leg, bent down, is the ancilla.

Based on this, we can write Uhlmann's theorem as formulated in Eqs.~\eqref{eq:uhlmann_definition_2} and~\eqref{eq:uhlmann_definition_3} for the case of MPSes as
\begin{align}
    \label{eq:uhlmann_MPS_1}
    F(\rho, \sigma) 
    &= \max_{W_{A}, W_{B}} \left| \; \includegraphics[scale=1,raise=-2.0em]{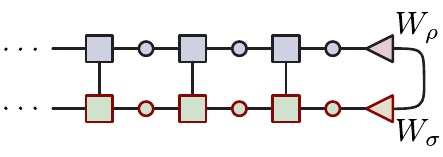} \; \right| \\%
    &= \;\, \max_{W} \;\,\, \left| \; \includegraphics[scale=1,raise=-2.0em]{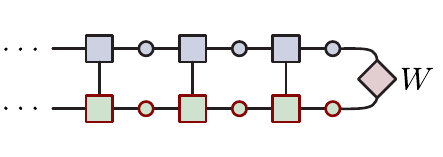} \; \right|.
    \label{eq:uhlmann_MPS_2}
\end{align}
Here the green tensors on the bottom row are the canonical form of the MPS $\ket{\phi}$, and together form $Y\dg$ of $\sigma = YY\dg$.
As discussed in Sect.~\ref{sec:fidelity_basics}, the optimal $W$ to maximize the expression in Eq.~\eqref{eq:uhlmann_MPS_2} is easily obtained from the singular value decomposition of the matrix
\begin{equation}
    \includegraphics[scale=1,raise=-2.0em]{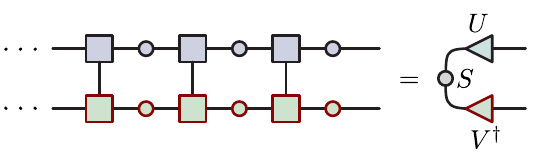} \;,
\end{equation}
as \includegraphics[scale=1, valign=c, raise=0.10em]{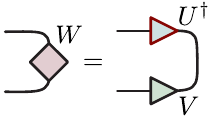}, which then yields
\begin{equation}
    F(\rho, \sigma) = \Tr S = \left\lVert \; \includegraphics[scale=1,raise=-2.0em]{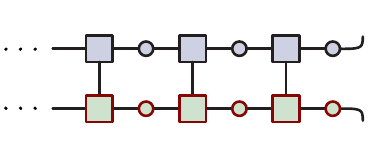} \; \right\rVert_{\text{tr}}.
    \label{eq:mixed_transfer_matrix}
\end{equation}
where we have bent the external legs of the matrix for which to trace norm is taken, for readability.
If the two MPSes have bond dimension $\chi_\psi$ and $\chi_\phi$, then this matrix is of dimensions $\chi_\psi \times \chi_\phi$.

We thus arrive at the following algorithm to evaluate the Uhlmann fidelity in Eq.~\eqref{eq:fidelity_definition}:
\begin{enumerate}
    \item Gauge transform the MPSes for $\ket{\psi}$ and $\ket{\phi}$ into the canonical form, shown in Eqs.~\eqref{eq:MPS_A_canonical} and~\eqref{eq:MPS_orthogonality}.\footnote{%
            In fact, transforming the whole MPS into the canonical form is not necessary.
            It is sufficient to only gauge transform the bond at $x$, such that it labels the Schmidt values and vectors of the left-right bipartition at $x$.
            }
    \item Construct the matrix of Eq.~\eqref{eq:mixed_transfer_matrix}.
    \item Evaluate the trace norm of this matrix. This norm is the Uhlmann fidelity $F(\rho, \sigma)$.
\end{enumerate}
The computational time cost of this procedure scales as $\mathcal{O}(\chi^3)$ for an MPS of bond dimension $\chi$, which is the same as the scaling of other typical MPS operations.


\subsubsection{Windows}%
\label{sec:window_fidelities_for_MPS}
Consider now the same setup as before, of two MPSes $\ket{\psi}$ and $\ket{\phi}$, but this time assume we want to evaluate the fidelity of their reduced density matrices $\rho$ and $\sigma$, not on half the system, but on a finite window in the middle.
We call this window $M$, and say that it is between two half-integer points on the lattice, $x_0$ and $x_1$.
We denote the parts of the lattice to the left and the right of $M$ by $L$ and $R$:
\begin{equation}
    \ket{\psi} = \includegraphics[scale=1,raise=-2.0em]{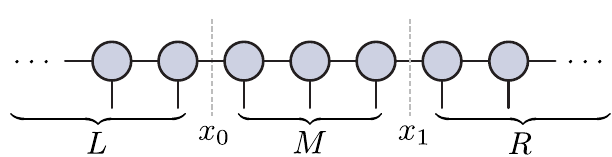} \;.
\end{equation}
As above in Sect.~\ref{sec:halfsystem_fidelities_for_mps}, we wish to use Uhlmann's theorem, and thus need the generic form of a purification $\ket{\varphi_\rho}$ of $\rho = \Tr_{M^\complement} \ket{\psi} \bra{\psi}$ (and similarly for $\ket{\phi}$ and $\sigma$).\footnote{%
    As in Sect.~\ref{sec:halfsystem_fidelities_for_mps} the MPS may be infinite or finite, as long as it does not have periodic boundaries.
    The periodic boundary condition case can also be treated, and at the same computational cost, but the procedure needs smalls changes due to the lack of a canonical form.
    }
Again we rely on the canonical form
\begin{align}
    \ket{\psi} = \; \includegraphics[scale=1, valign=c,raise=-0.2em]{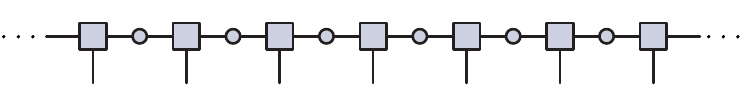} \;,
\end{align}
and using its orthogonality properties from Eq.~\eqref{eq:MPS_orthogonality}, we obtain
\begin{align}
    \rho &= \; \includegraphics[scale=1, valign=c, raise=-0.2em]{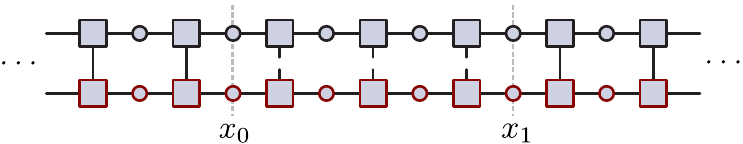}\\
    &= \quad \qquad \qquad \; \includegraphics[scale=1, valign=c]{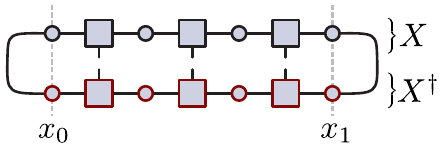} \;,
\end{align}
which is of the form $\rho = XX\dg$ that we need to make use of the results in Sect.~\ref{sec:fidelity_basics}.
Based on this we know that the generic form of a purification of $\rho$ is
\begin{align}
    \includegraphics[scale=1, valign=c]{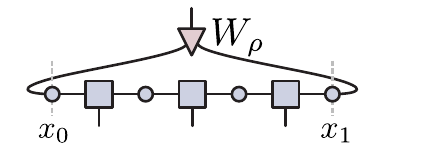} \;,
\end{align}
and that the fidelity of $\ket{\psi}$ and $\ket{\phi}$ on $M$ is
\begin{align}
    \label{eq:window_fidelity_MPS_1}
    F(\rho, \sigma) &= \max_W \left| \; \includegraphics[scale=1, valign=c]{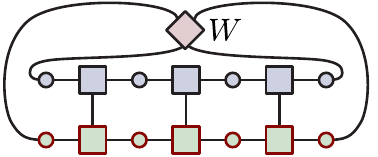} \; \right|\\%
    &= \qquad \left\lVert \;\, \includegraphics[scale=1, raise=-1.2em]{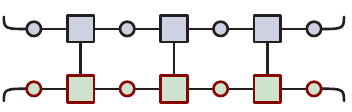} \;\, \right\rVert_{\text{tr}}\;,
    \label{eq:window_fidelity_MPS_2}
\end{align}
where again $W$ is constrained to be isometric between the top and the bottom legs, and thus the solution to the maximization problem is the trace norm of the transfer matrix in Eq.~\eqref{eq:window_fidelity_MPS_2}, when viewed as matrix between the top two and the bottom two legs.

Eq.~\eqref{eq:window_fidelity_MPS_1} can be a useful quantity to evaluate, but computing it does require $\mathcal{O}(\chi^6)$ time, compared to all the usual MPS operations, which can be done in $\mathcal{O}(\chi^3)$ time.
(Notice that for a periodic MPS, $\mathcal{O}(\chi^6)$ is the usual leading order cost~\cite{zou_conformal_2017}.)
This is because the isometry $W$ is a $\chi^2 \times \chi^2$ matrix\footnote{%
    Or more generally $\chi_\psi^2 \times \chi_\phi^2$, if the two MPSes have different bond dimensions.
    }, that connects both ends of the region $M$.
$W$ is answering the question ``How large can the overlap of the two states be, if outside of $M$ they are allowed to match each other perfectly?'', and it is answering it in a way that allows the two, disconnected ends of the system, $L$ and $R$, to conspire with each other.
One natural question to ask is, whether it is necessary for the left and the right ends to be correlated to maximize this overlap.
This can be answered by doing the maximization of Eq.~\eqref{eq:window_fidelity_MPS_1}, but with the restriction that $W$ is a tensor product of two disjoint isometries at the two ends:
\begin{equation}
    \label{eq:separate_window_fidelity_MPS}
    \Fsep (\rho, \sigma) = \max_{W_L, W_R} \left| \; \includegraphics[scale=1, valign=c]{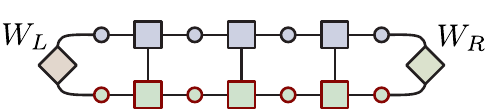} \right|,
\end{equation}
with $W_L$ and $W_R$ isometric.
The optimal choice of $W_L$ and $W_R$ is no longer a single straight-forward SVD, since choice of one affects the other.
However, the optimization can be done by holding one of $W_L$ and $W_R$ fixed while optimizing the other as in Eq.~\eqref{eq:uhlmann_MPS_2}, and repeating this procedure, alternating between $W_L$ and $W_R$ until convergence is reached.
Each iteration can be done in $\mathcal{O}(\chi^3)$ time, and we find that the process usually converges in just a few iterations.

This new measure of fidelity over $M$ from Eq.~\eqref{eq:separate_window_fidelity_MPS}, which we call $\Fsep$ or \emph{disjoint fidelity}, can be compared to the usual Uhlmann fidelity $F$.
First, note that since the product $W_L \otimes W_R$ is a valid choice for the isometry $W$ in Eq~\eqref{eq:window_fidelity_MPS_1}, $\Fsep$ is a strict lower bound for $F$.
Second, if the region $M$ is larger than the correlation lengths of the MPSes, then $\Fsep \approx F$, as one end of $M$ is essentially uncorrelated from the other.
$\Fsep$ can be seen either as a cheap and conservative approximation to $F$, or as a separate notion of fidelity, that forbids collusion between disjoint parts of $M^\complement$ in the purification.


\subsection{Tree Tensor Networks}%
\label{sec:fidelities_ttns}
In this section, we concentrate on Tree Tensor Networks, or TTNs.
Like MPSes, they too can be used to represent low entanglement states of 1D lattice systems.
They naturally support entanglement structures that resemble a tree, and implement a notion of coarse-graining for lattice systems.
More background about TTNs can be found for instance in Refs.~\onlinecite{shi_classical_2006,tagliacozzo_simulation_2009,2009arXiv0912.1651V}.

A TTN is a tensor network of the following form:
\begin{equation}
    \label{eq:ttn}
    \includegraphics[scale=1, valign=c]{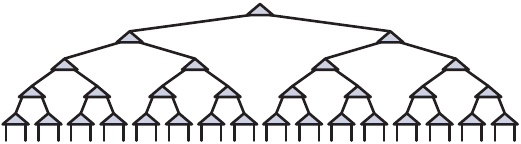}.
\end{equation}
The open indices at the bottom are the physical lattice sites, and for simplicity of discussion we assume that all the contracted indices have bond dimension $\chi$.
The tensors in a TTN are constrained to be isometric in the sense that
\begin{equation}
    \label{eq:ttn_isometricity}
    \includegraphics[scale=1, valign=c]{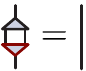} \;.
\end{equation}
Written in the traditional linear algebra notation, if $w$, of dimensions $\chi \times \chi^2$, is the tensor of the TTN, then the isometricity condition is $ww\dg = \unity$.

As with MPSes, certain subsystem fidelities are more natural and efficient to compute for TTNs than others.
The characterizing criterion is, how many indices need to be cut to be able to separate a given subsystem from its complement.
For MPSes, left-right bipartitions can be done by cutting only one index, and thus evaluating subsystem fidelities for them was simple and computationally cheap.
Similarly for a TTN, the subsystems that can be separated from the rest by cutting a single leg allow for computing the fidelity with the lowest effort and computational cost.
These subsystems are finite windows of size $2^n$, that correspond to branches of the tree.
This means every single-site subsystem, every second contiguous two-site block, every fourth contiguous four-site block, etc.
Below are shown some examples of such subsystems, underlined in red, together with the single-leg cuts that separate them from their complements.
\begin{equation}
    \label{eq:ttn_braces}
    \includegraphics[scale=1, valign=c]{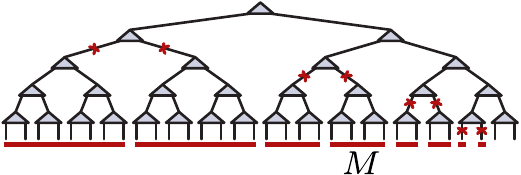}.
\end{equation}

As an example, let us show how to compute the subsystem fidelity between two TTN states on the subsystem marked above as $M$.
Call the state in Eq.~\eqref{eq:ttn_braces} $\ket{\psi}$, and the reduced density matrix $\rho = \Tr_{M^\complement} \ket{\psi}\bra{\psi}$.
Using the isometricity condition~\eqref{eq:ttn_isometricity}, it is easy to see that
\begin{equation}
    \label{eq:ttn_rho_1}
    \rho \; = \; \includegraphics[scale=1, valign=c]{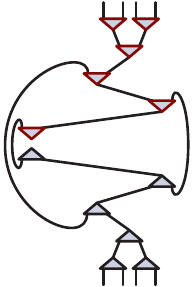},
\end{equation}
where again red boundaries on tensors mark complex conjugation.
Eq.~\eqref{eq:ttn_rho_1} is already of the form $\rho = XX\dg$ that we need, but $X$ has a very large number of columns, namely $\chi^n$, with $n$ being the number of vertical legs passing through the middle of the diagram, in this case $3$.
To improve the situation, we contract the middle part of the diagram in Eq.~\eqref{eq:ttn_rho_1} and then decompose it:
\begin{equation}
    \label{eq:ttn_rho_2}
    \rho \; = \; \includegraphics[scale=1, valign=c]{ttn_rho_1} \; = \; \includegraphics[scale=1, valign=c]{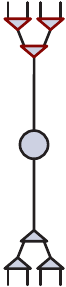} \; = \includegraphics[scale=1, valign=c]{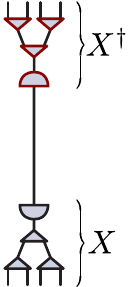} \;.
\end{equation}
At the final step, the decomposition of the round matrix in the middle uses its positive semidefiniteness.

Eq.~\eqref{eq:ttn_rho_2} is of the $\rho = XX\dg$ form, but with $X$ now having only $\chi$ columns, which makes it computationally manageable.
From this point on we can invoke Eqs.~\eqref{eq:uhlmann_definition_2} and~\eqref{eq:uhlmann_definition_3} as we did with MPSes, and arrive at the following expression for the Uhlmann fidelity of $\rho = \Tr_{M^\complement} \ket{\psi} \bra{\psi}$ and $\sigma = \Tr_{M^\complement} \ket{\phi} \bra{\phi}$:
\begin{align}
    \label{eq:ttn_fidelity_1}
    F(\rho, \sigma) 
    \;=\; \max_{W} \left| \; \includegraphics[scale=1,raise=-2.4em]{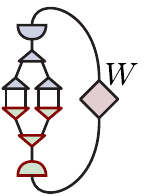} \right|
    \;=\; \left\lVert \; \includegraphics[scale=1,raise=-2.4em]{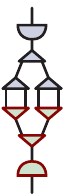} \; \right\rVert_{\text{tr}}.
\end{align}
Here, again, $W$ is an isometry, and the green tensors are the tensors from the TTN $\ket{\phi}$, whereas the blue ones are from $\ket{\psi}$.

Constructing the matrix in Eq.~\eqref{eq:ttn_fidelity_1} and evaluating its trace norm can be done in $\mathcal{O}(\chi^4 \log_2 L)$ time, with $L$ being the system size and $\chi$ the bond dimension\footnote{%
    We assume again for simplicity that $\ket{\psi}$ and $\ket{\phi}$ have the same bond dimension.
    This need not be the case.
    }.
Since most TTN operations necessary to optimize such a state or evaluate observables from it scale as $\mathcal{O}(\chi^4 \log_2 L)$ or worse, evaluating these subsystem fidelities is never the bottleneck of the computation.
Although we presented here how to evaluate fidelities for the subsystem $M$ from Eq.~\eqref{eq:ttn_braces}, the same procedure applies to any subsystem that can be separated by a single cut.

As in the case of MPSes, fidelities for other subsystems can also be evaluated, although typically at higher computationally cost.
Similar notions of disjoint fidelity as the one in Sect.~\ref{sec:window_fidelities_for_MPS} can also be defined, by restricting the maximization in Uhlmann's theorem to purifications that limit correlations between disjoint parts of $M^\complement$.
We omit the general analysis due to its complexity, but specific choices of $M$ can easily be analysed case-by-case.

Here we have concentrated on TTNs as they are most commonly used in many-body physics, with the isometricity constraint~\eqref{eq:ttn_isometricity}.
Consider now relaxing the isometricity condition, and furthermore allowing the graph of contractions for the tensor network to be any tree, as opposed to the binary trees of fixed depth discussed above.
This larger class of tensor networks is exactly that of acyclic graphs, meaning networks that have no closed loops.
Again we can consider subsystems that can be separated by cutting a single index in the network, and the above analysis requires slight modifications, but the result remains the same:
Subsystem fidelities for these subsystems, between two states that have the same tree-graph of contractions, can be evaluated efficiently and easily.
The computational cost scales with a power of the bond dimension $\chi$, that is the same as for most operations needed for the tensor networks in question (for instance, for a ternary tree, most basic operations scale as $\mathcal{O}(\chi^5)$ in $\chi$).
For more details on how to implement this for a generic tree, see Ref.~\onlinecite{shi_classical_2006}.


\section{Applications}%
\label{sec:applications}
The ability to evaluate Uhlmann fidelities for subsystems allows us to spatially resolve the overlaps of pure states.
In this section we give some example applications of where this is useful.
When a benchmark model is needed, we use the 1D Ising model with a transverse field:
\begin{equation}
    \label{eq:ising_hamiltonian}
    H_{\text{Ising}} = -\frac{1}{2} \sum_i \left( X_i X_{i+1} + h Z_i - \frac{4}{2\pi} \unity \right).
\end{equation}
The external magnetic field $h$, chosen to be $h \ge 0$, distinguishes two phases, a symmetry breaking one for $h < 1$ and a disordered one for $h > 1$, which are separated by a critical point at $h=1$.
The normalization in the Hamiltonian~\eqref{eq:ising_hamiltonian} has been chosen such that the ground state energy is $0$ and the slope of the dispersion relation at low energies is $1$.

\subsection{Local quench}%
\label{sec:local_quench}
Consider a Hamiltonian $H$ and its ground state $\ket{E_0}$.
We may ask what happens in the time-evolution after a local quench, where the state is perturbed with some local operator $O_x$ at site $x$, and then time evolved by time $t$, to reach the state $\ket{\psi(t)} = e^{itH} O_x \ket{E_0}$.
Presumably the effect of the perturbation has spread to a region around $x$, and we may ask for instance, where is most of the perturbation ``located'', and has some part of the system returned to its original state.
These questions can be answered by computing subsystem fidelities between $\ket{\psi(t)}$ and the unperturbed $\ket{E_0}$.

\begin{figure}
    \includegraphics[width=\linewidth]{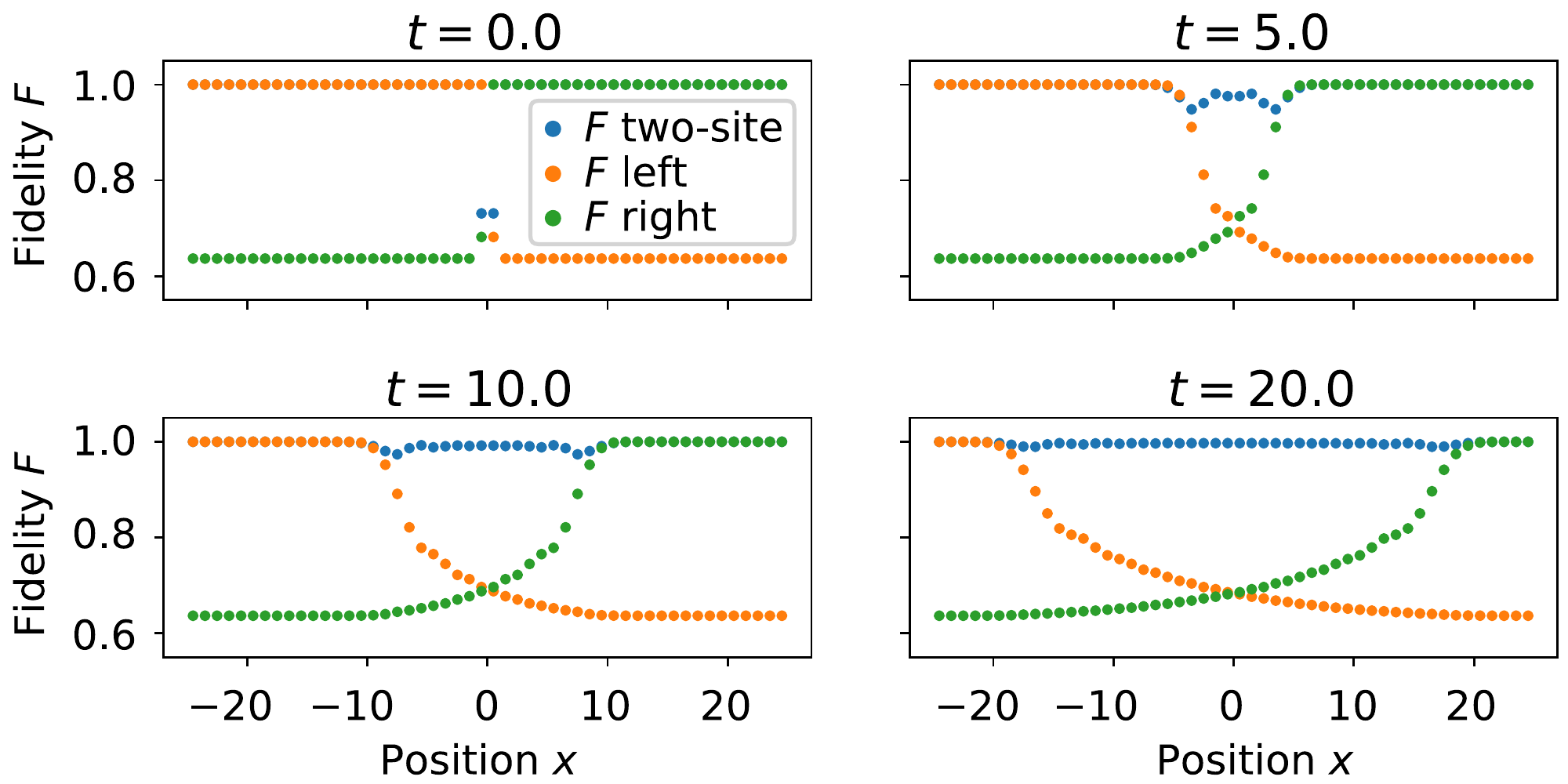}
    \caption{%
        Subsystem fidelities between the ground state $\ket{E_0}$ of the infinite, critical Ising model, and the locally quenched state $e^{itH_{\text{Ising}}} Z_0 \ket{E_0}$, as functions of position $x$, at various times $t$ after the quench.
        The three different fidelities plotted are the half-system Uhlmann fidelities to the left (descending orange dots) and to the right (ascending green dots) of $x$, and the two-site fidelity at $x$ (blue dots).
        A bond dimension 50 MPS was used in the time evolution.
        }
    \label{fig:local_quench}
\end{figure}

To illustrate the idea, we perturb the ground state of the infinite, critical Ising model, represented as an MPS, with the Pauli $Z$ operator, and time evolve to obtain $\ket{\psi(t)} = e^{itH_{\text{Ising}}} Z_0 \ket{E_0}$, where we have chosen to call the position of the $Z$ insertion the origin.
For various positions $x$, ranging from $x \ll t$ to $t \ll x$, we then compute three different fidelities between $\ket{\psi(t)}$ and $\ket{E_0}$:
The window fidelity for a two-site window around $x$; the half-system fidelity for the system left of $x$; and a similar half-system fidelity but for the right.
These fidelities, evaluated at various times $t$, are shown in Fig.~\ref{fig:local_quench}, and one of them, for $t=10$, was already used as an example in the introduction.

Several observations can be made from these results.
To start off, as a sanity check, it is good notice that the half-system fidelities start from $1$, since far away from the perturbation its effect is not seen, then decay monotonously as the size of the subsystem that they are computed on increases, and finally asymptote to the expectation value $|\braket{\psi(t)}{E_0}| = |\bra{E_0} Z \ket{E_0}|$.
Next, note that with the normalization of the Hamiltonian chosen in Eq.~\eqref{eq:ising_hamiltonian}, the ballistic front propagates at speed 1, and correspondingly we see that the time evolved state $\ket{\psi(t)}$ differs from the unperturbed state $\ket{E_0}$ most strongly at the fronts $x \approx t$ and $x \approx -t$.
In the region $-t \ll x \ll t$, where the propagation of the perturbation has already ``passed by'', the two-site fidelity reports that the state mostly looks like the ground state, but the half-system fidelities keep decreasing.
Finally, notice an interesting asymmetry in the behavior of the half-system fidelities:
They show a sharp decline when they meet the first ballistic front, but the final dip down to the asymptotic value at the second front is only a very small one.

\subsection{Comparing states at different scales}%
\label{sec:comparing_phases}
Another instance where spatially resolving the overlap between two states is of interest are cases where the states are translation invariant, and similar at some length scales, while different at others.
One such circumstance is comparing ground states of the same model at slightly different values of the couplings.
Such ground state fidelities are useful to explore many-body physics, including first order and continuous phase transitions, as discussed in Ref.~\onlinecite{zhou_ground_2008}.
As an example, we again consider the Ising Hamiltonian~\eqref{eq:ising_hamiltonian} on an infinite system.
We take its ground states at the critical point $h=1.0$ and slightly in the disordered phase at $h=1.01$, represented as MPSes, and compare their fidelities over finite windows of varying sizes.

\begin{figure}
    \includegraphics[width=\linewidth]{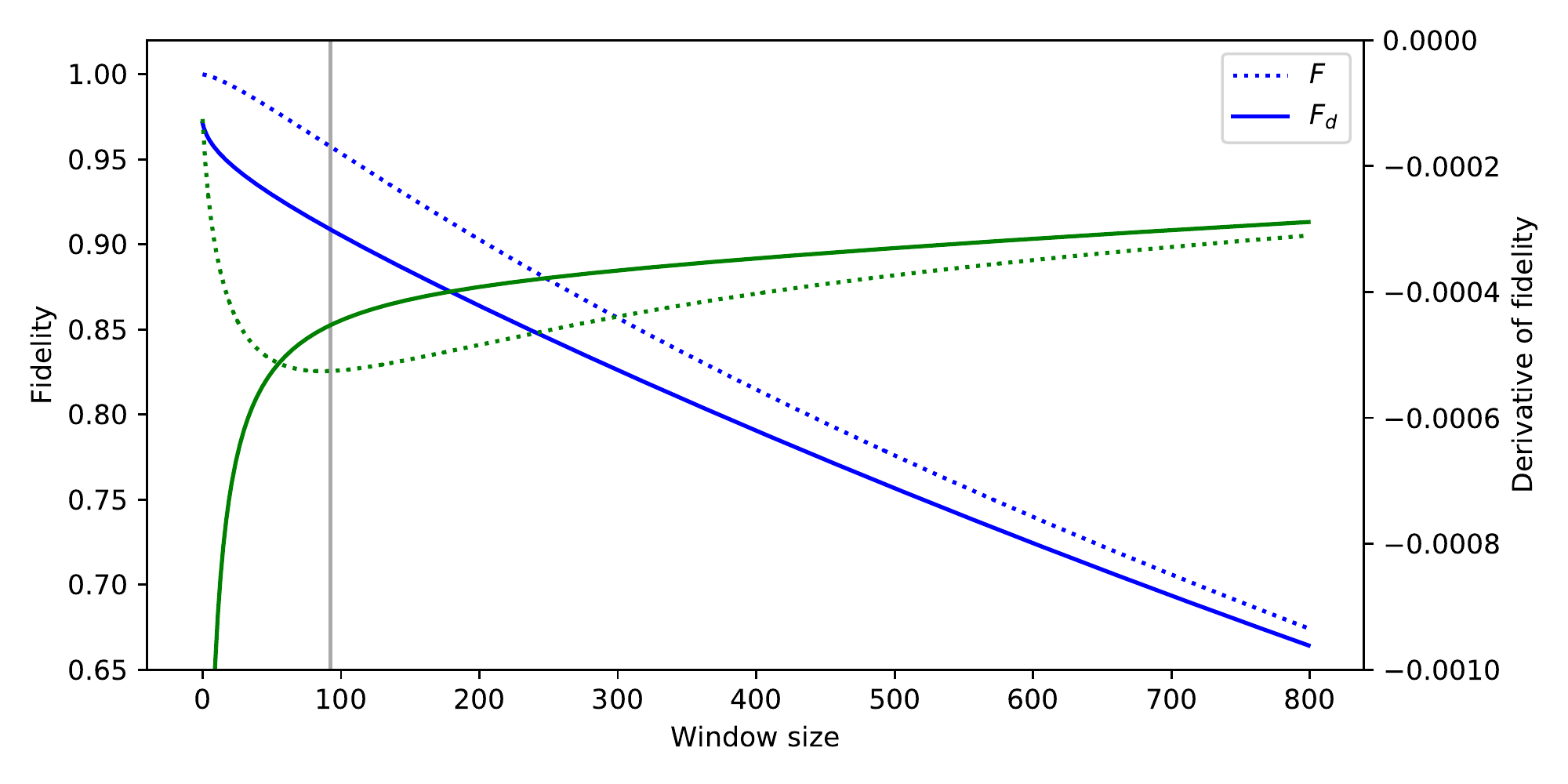}
    \caption{%
        Window fidelities between the ground states of the infinite, critical Ising model at the critical point $h=1.0$ and slightly in the disordered phase $h=1.01$, as a function of the size of the window.
        The monotonously decreasing blue lines are the fidelities, dotted line for the Uhlmann fidelity $F$ and solid line for the disjoint fidelity $\Fsep$, with their axis on the left.
        The green lines are the discrete derivatives of the blue lines, with their axis on the right.
        The vertical grey line marks the correlation length of the $h=1.01$ ground state.
        Bond dimension 50 MPSes were used to generate these results.
        }
    \label{fig:comparing_phases}
\end{figure}

Results are shown in Fig.~\ref{fig:comparing_phases}.
The most immediately visible feature is the disagreement between the Uhlmann fidelity (dotted blue line) and the disjoint fidelity (solid blue line).
This is a consequence of the long-ranged nature of the critical state, where correlations persist over all length scales.
This means that the optimization for the isometry $W$ in Eq.~\eqref{eq:window_fidelity_MPS_1} benefits from being able to bridge the two ends of the window, compared to the disjoint fidelity that forbids this.
The two fidelities get closer to each other at larger distances.

Let us now concentrate on the Uhlmann fidelity, as it is a more sensitive measure of the similarity of the two states.
Intuition based on the renormalization group would suggest that the almost-critical state at $h=1.01$ should look mostly like the critical one at short length scales, and then significantly differ at long length scales.
This behavior can be seen in the Uhlmann fidelity (dotted blue line) in Fig.~\ref{fig:comparing_phases}:
The slope of the curve starts mostly flat, and then dips down around the correlation length of the off-critical state, marked with the grey vertical line, before settling into a steady exponential decay.
This feature is easier to see in the derivative $\frac{\partial F}{\partial |M|}$, which is plotted with the green dotted line, and shows a minimum close to the correlation length.
In other words, adding one more site to the window $M$ causes the largest change in the fidelity when the size of $M$ is close to the correlation length, demonstrating that the difference between the critical and the off-critical states is the most pronounced at these length scales.
In results not shown here, we observe that the minimum of the derivative follows the correlation length of the off-critical state for a wide range of values for $h$.

\subsection{Convergence of simulations}%
\label{sec:convergence}
As with most numerical methods, tensor network algorithms typically require iterative optimizations, and have parameters that control the level of approximations, namely the bond dimensions.
When simulating a given system, one needs to then ask, has the optimization converged, and were the bond dimensions used large enough to faithfully describe the physics.
In this section, we demonstrate using fidelities $F$ and $\Fsep$ to answer these kinds of questions.

Consider two MPSes $\ket{\psi}$ and $\ket{\phi}$, with different bond dimensions $\chi_\psi$ and $\chi_\phi$, that have both been optimized to minimize the energy for a given Hamiltonian $H$.\footnote{%
    Many different optimization algorithms could be used.
    We choose here to use imaginary time evolution, implemented using a Matrix Product Operator representation of $e^{-\tau H}$.
}
Especially if $H$ is critical or nearly critical, we may worry that the bond dimensions we have chosen may not be sufficient to accurately represent the ground state of $H$.
For critical systems, we in fact know that no finite bond dimension is sufficient to describe the long distance physics correctly, but we would still hope that for distances shorter than the effective correlation lengths of $\ket{\psi}$ and $\ket{\phi}$ (imposed by the finite bond dimension), the MPSes would approximate the ground state well.
To test whether our hopes are fulfilled or our bond dimensions are too small, we can compute $F$ or $\Fsep$ of $\ket{\psi}$ and $\ket{\phi}$, for finite windows of various sizes:
If for a window of size $|M|$, the subsystem fidelity of $\ket{\psi}$ and $\ket{\phi}$ is far from one, the simulations can not be trusted to faithfully represent the physics at distances of order $|M|$.

\begin{figure}
    \includegraphics[width=\linewidth]{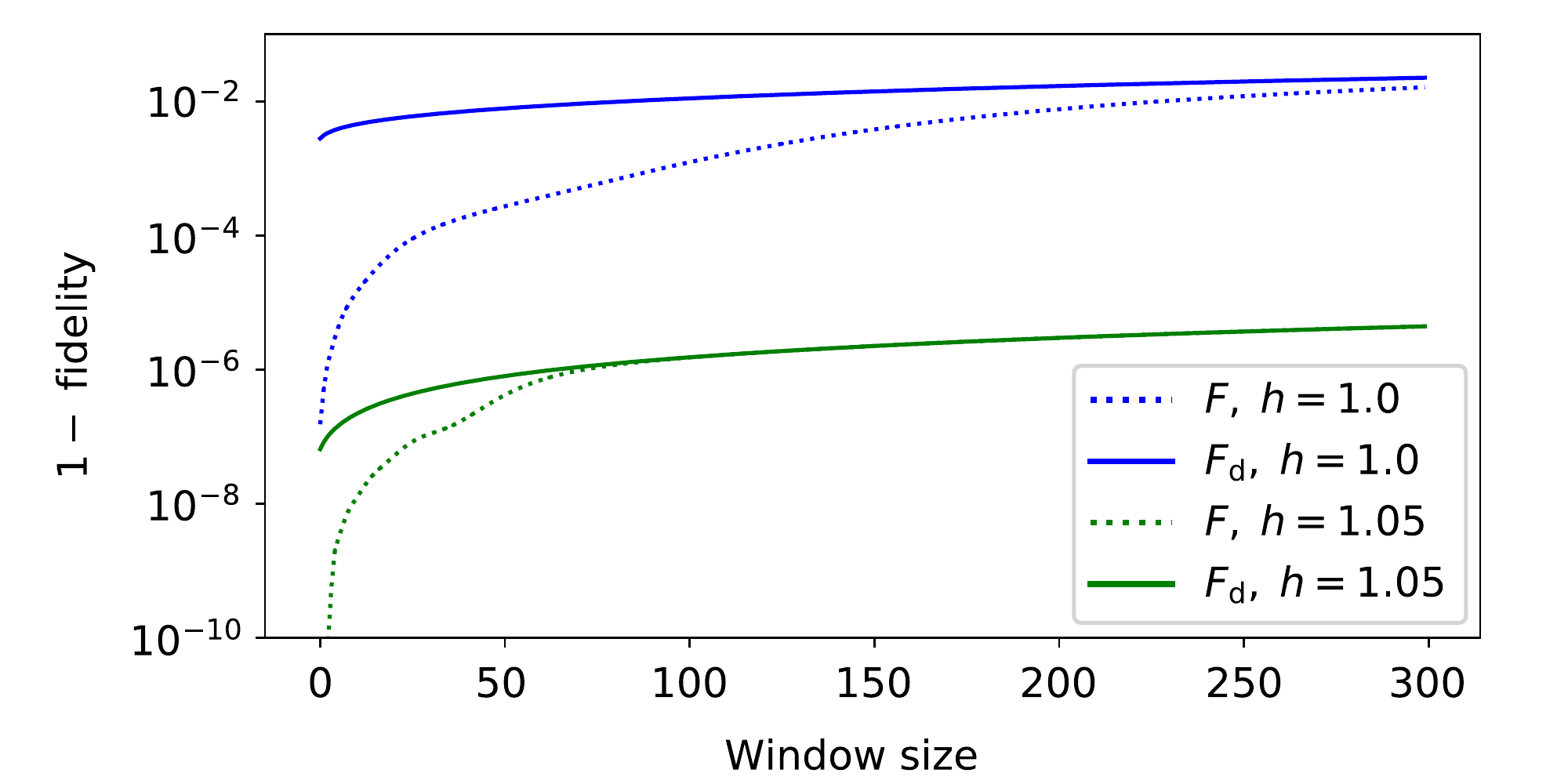}
    \caption{%
        Window fidelities of pairs of MPSes, representing the same ground state, but with different bond dimensions 10 and 20, as functions of the window size $|M|$.
        In the optimal case these fidelities would be exactly one, so the vertical axis is the difference $1 - \text{fidelity}$, on a logarithmic scale.
        Dotted lines mark the Uhlmann fidelities $F$, solid lines are disjoint fidelities $\Fsep$.
        In both cases the model is the infinite Ising model, with the blue lines at the top being for ground states of the $h=1.0$ critical Hamiltonian, and the green ones at the bottom for $h=1.05$.
        }
    \label{fig:convergence_in_chi}
\end{figure}

In Fig.~\ref{fig:convergence_in_chi} we benchmark this idea, using again the Ising model.
On its vertical axis is $1 - \text{fidelity}$, where fidelity is the Uhlmann fidelity $F$ for the dotted lines and the disjoint fidelity $\Fsep$ for the solid lines.
Each line in the figure is the fidelity between two MPS approximations to the same state, at bond dimensions 10 and 20.

The green lines show fidelities at $h=1.05$.
Both the Uhlmann fidelity and the disjoint fidelity remain very close to one, which signals that bond dimension 10 is probably already sufficient for accurately describing the state, at least up to distances of 300 sites.
At short distances the disjoint fidelity significantly underestimates the Uhlmann fidelity, but it is always a lower bound for it, and for high bond dimensions would be much faster to compute.

The blue lines show a similar comparison, but now at the critical point $h=1.0$.
The more entangled nature of the ground state makes it harder for the MPS to faithfully represent the state, which shows as a large difference between the $\chi_\psi = 10$ and $\chi_\phi = 20$ states, calling any long-range properties evaluated from these MPSes into question.
The disjoint fidelity is seen to much more grossly underestimate the Uhlmann fidelity at short distances, due to the long-range correlations in the state.
Finally, note that at short distances the Uhlmann fidelity remains quite large, which means that up to a distance of a few dozen sites, the state already has converged in bond dimension to a reasonable accuracy.

\begin{figure}
    \includegraphics[width=\linewidth]{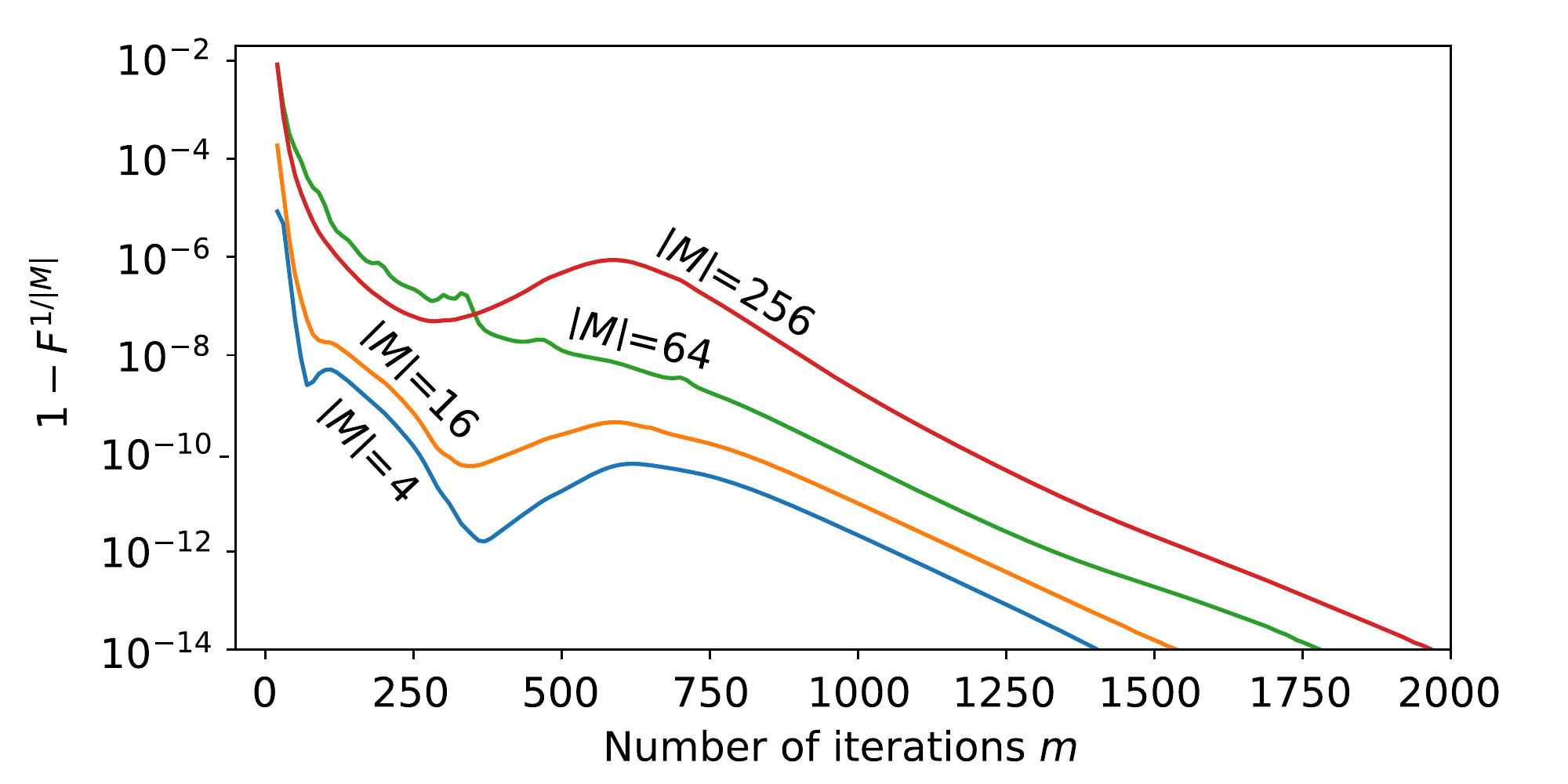}
    \caption{%
        Per-site window fidelities $F^{\frac{1}{|M|}}$ of two TTN states $\ket{\psi_m}$ and $\ket{\psi_{m-10}}$ as a function of the number of iterations $m$ in an optimization algorithm.
        Different lines correspond to different window sizes $|M|$, which furthermore relate to different layers of the TTN\@.
        The TTN used here consists of 8 layers and is enforced to be translation and reflection invariant, meaning that the position of the window does not matter.
        The optimization is for the ground state of the critical Ising model, and the bond dimension of the TTN is 30.
        }
    \label{fig:ttn_convergence}
\end{figure}

This kind of analysis can be done not only for convergence in bond dimension, but also for convergence during an iterative optimization.
As an example of this, we consider a Tree Tensor Network, that is iteratively optimized to minimize its energy with respect to the critical Ising Hamiltonian.
Let us denote by $\ket{\psi_m}$ the TTN state that has gone through $m$ iterations of the optimization algorithm.
The optimization starts from a random TTN at $\ket{\psi_0}$, and eventually at $\lim_{m \to \infty} \ket{\psi_m}$ converges to the best possible approximation to the Ising ground state that our chosen bond dimension allows.\footnote{%
    This is assuming the optimization does not get stuck in a local minimum.
    }
In Fig.~\ref{fig:ttn_convergence}, we use Eq.~\eqref{eq:ttn_fidelity_1} to compute window fidelities between the states $\ket{\psi_m}$ and $\ket{\psi_{m-10}}$, and plot the results as a function of $m$.
This provides a measure of convergence, since the fidelity of $\ket{\psi_m}$ and $\ket{\psi_{m-10}}$ tells us how much the state has changed over the last 10 iterations.
Results are shown separately for various window sizes $|M|$, and to provide a fair comparison of fidelities at different $|M|$, we look at the per-site fidelity $F^{\frac{1}{|M|}}$.
This allows us to observe that for small window sizes the state seems to converge relatively fast, compared to larger windows which keep changing significantly for many hundreds of iterations.
Since in a TTN the lower layers of the tree dictate the short distance properties of the state, and conversely the higher layers relate to long distances, we conclude that the layers converge at different speeds, with the lower layers converging first, followed then by the higher ones.
Resurgences can also be seen during the optimization, where some layers that have already converged to quite a high accuracy, suddenly change significantly, due to the highly non-linear nature of the optimization.
Note that convergence of the state, as witnessed by the fidelity, is a much more stringent criterion than convergence in any single observable, such as the energy.

\section{Conclusion}%
\label{sec:discussion}
In this paper we explain how to compute subsystem fidelities for many-body systems using tensor network states.
Such fidelities give a spatial characterization of differences between two states, that is agnostic about the nature of the degrees of freedom or the interactions.
We demonstrate their usefulness with example applications:
We study a local quench, resolve in scale the difference between a critical and an off-critical state, and use similarity between states as a measure of convergence in a simulation.

Other applications, not discussed here, are also possible~\cite{liu_local_2017,banchi_quantum_2014}.
For instance, one could study the effect of an impurity in the Hamiltonian, by comparing low-energy states with and without the purity.
One could also study the bipartite entanglement between two halves of a system, and characterize it beyond the entanglement spectrum, by resolving the orthogonality of the Schmidt vectors:
The Schmidt vectors are by definition orthogonal to each other, but their subsystems fidelities may decay in different ways as functions of the size of the subsystem, informing us of how different parts of the system contribute to the entanglement.
We leave these, and possible other applications, for future study.

In Sect.~\ref{sec:evaluating_fidelities} we discussed how to evaluate subsystem fidelities for Matrix Product States and Tree Tensor Networks.
The reason we concentrated on these two network types is that they allow for separating certain subsystems from their complements by cutting only one or two indices.
This means that the corresponding reduced density matrices have small-rank decompositions of the form $\rho = XX\dg$, which allowed us to evaluate Uhlmann fidelities at relatively low cost.
It is worth pointing out that for some other networks, such as MERA~\cite{vidal_class_2008} or PEPS~\cite{nishio_tps_2004,Verstraete:2004cf}, this is not the case.
For instance, in a MERA, separating a region of length $L$ requires cutting $n \sim \mathcal{O}(\log L)$ indices, whereas in a PEPS separating a region of $L \times L$ requires cutting $n \sim \mathcal{O}(L)$ indices.
The rank of $\rho$ is exponential in $n$, and so is the cost of computing the Uhlmann fidelities.
Thus for both of these networks, evaluating subsystem fidelities is only feasible for relatively small subsystems.

Finally, note that both MPS and TTN are useful ansätze for 2-dimensional systems too~\cite{2011arXiv1105.1374S,tagliacozzo_simulation_2009,yan_spinliquid_2011,cincio_characterizing_2012,jiang_identifying_2012,depenbrock_nature_2012}, and the methods we describe can be applied in that context as well.

\vspace{1cm}

\emph{Note added.}
After the publication of the first version of this paper on the arXiv, we became aware of related recent work by Liu, Gu, Li, and Wang, in Ref.~\onlinecite{liu_local_2017}.
In its appendix, the authors discuss how to evaluate half-system fidelities for MPSes, reaching the same result as we do in Sect.~\ref{sec:halfsystem_fidelities_for_mps}, as well as discussing a subsystem not covered in this paper.
Here we present how subsystem fidelities can be evaluated from tensor networks more generally, giving a more detailed analysis of the MPS case, as well as extending the discussion to Tree Tensor Networks.
The main text of Ref.~\onlinecite{liu_local_2017} concentrates on using subsystem fidelities to identify and locate zero-modes in symmetry protected topologically ordered states and discrete symmetry breaking states, providing an interesting application of subsystem fidelities, and complementing the example applications we discuss.


\begin{acknowledgments}
The authors thank Stefan Kühn for providing the Tree Tensor Networks analyzed in Fig.~\ref{fig:ttn_convergence}.
The authors acknowledge support from the Simons Foundation (Many Electron Collaboration).
M.~Hauru is supported by an Ontario Trillium Scholarship.
G.~Vidal acknowledges support by a Natural Sciences and Engineering Research Council of Canada (NSERC) Discovery Grant.
Computations were made on the supercomputer Mammouth Parall{\`e}le~2 from the Universit{\'e} de Sherbrooke, managed by Calcul Qu{\'e}bec and Compute Canada.
The operation of this supercomputer is funded by the Canada Foundation for Innovation (CFI), the minist{\`e}re de l'{\'E}conomie, de la science et de l'innovation du Qu{\'e}bec (MESI) and the Fonds de recherche du Qu{\'e}bec -- Nature et technologies (FRQ-NT).
This research was supported in part by Perimeter Institute for Theoretical Physics.
Research at Perimeter Institute is supported by the Government of Canada through the Department of Innovation, Science and Economic Development Canada and by the Province of Ontario through the Ministry of Research, Innovation and Science.
\end{acknowledgments}

\nocite{5725236,SciPy}

\newpage
\appendix

\section{Generic form of purifications}%
\label{app:generic_purification}
In this appendix, we prove the following theorem, that characterizes all purifications of a given density matrix.

\begin{thrm:generic_purification}
    \label{thrm:generic_purification}
    Let $\rho$ be a density matrix on the $\chi$-dimensional state space $\statespace$.
    Let $X$ be a $\chi \times \chi_X$ matrix such that $\rho = XX\dg$.
    Let $\ket{\varphi} \in \statespace \otimes \ancillaspace$ be a candidate for being a purification of $\rho$.
    $\ancillaspace$ is an ancilla space of dimension $\chi_{\varphi} \ge \chi_X$.
    Let $\varphi$ be a $\chi \times \chi_{\varphi}$ matrix dual to the state $\ket{\varphi}$, in the sense that $\bra{i} \varphi \ket{j} = [\bra{i} \otimes \bra{j}] \ket{\varphi}$ for all $\ket{i} \in \statespace$ and $\ket{j} \in \ancillaspace$.
    Then $\ket{\varphi}$ is a purification of $\rho$ if and only if there exists a matrix $W$ of dimensions $\chi_X \times \chi_{\varphi}$, such that
    \begin{itemize}
        \item $W$ is isometric in the sense that $W W\dg = \unity$,
        \item $\varphi = XW$.
    \end{itemize}
\end{thrm:generic_purification}

\begin{proof}
    First, assume $\varphi = X W$, with $W$ being isometric.
    Then
    \begin{equation}
        \Tr_{\text{anc.}} \ket{\varphi}\bra{\varphi} = \varphi \varphi\dg = XWW\dg X\dg = XX\dg = \rho,
    \end{equation}
    and thus $\ket{\varphi}$ is a purification of $\rho$.

    To prove the inverse statement, assume $\ket{\varphi}$ is a purification of $\rho$.
    Let $X = U_X S_X V_X\dg$ and $\varphi = U_\varphi S_\varphi V_\varphi\dg$ be the singular value decompositions of $X$ and $\varphi$.
    Then
    \begin{equation}
        \rho = XX\dg = U_X S_X S_X\dg U_X\dg
    \end{equation}
    and
    \begin{equation}
        \rho = \Tr_{\text{anc.}} \ket{\varphi}\bra{\varphi} = \varphi \varphi\dg = U_\varphi S_\varphi S_\varphi\dg U_\varphi\dg.
    \end{equation}
    $S_X S_X\dg$ and $S_\varphi S_\varphi\dg$ are square and diagonal, and $U_X$ and $U_\varphi$ are unitary, and thus the above are both eigenvalue decompositions of $\rho$.
    Furthermore, we can choose $S_X$ and $S_\varphi$ to have the singular values ordered by magnitude, which then makes $S_X S_X\dg$ and $S_\varphi S_\varphi\dg$ be ordered by magnitude.
    Since the eigenvalue decomposition of a Hermitian matrix is unique up to unitaries that commute with the diagonal matrix of eigenvalues, we then know that $S_\varphi S_\varphi\dg = S_X S_X\dg = S^2$ and $U_\varphi = U_X u$, where $u$ commutes with $S^2$.

    $S_\varphi$ and $S_X$ are in general non-square, and their right-most dimensions, $\chi_{\varphi}$ and $\chi_X$, may be different.
    However, since they both square to $S^2$ in the above sense, we know that the values on their diagonals must be the same.
    Thus the only difference between them is that  $S_\varphi$ may be padded with columns of zeros compared to $S_X$ (keep in mind that $\chi_\varphi \ge \chi_X$).
    We can in fact write $S_\varphi = S_X E$, with $E$ (for embedding) being the $\chi_X \times \chi_{\varphi}$ matrix
    \begin{equation}
        E =
        \begin{bmatrix}
            && & &&\\
            &\unity_{\chi_X \times \chi_X}& & &\bar{0}& \\
            && & &&
        \end{bmatrix},
    \end{equation}
    where $\bar{0}$ is zero matrix of dimensions $\chi_X \times (\chi_\varphi - \chi_X)$.

    For $u$, the fact that it commutes with $S^2$ implies that there exist unitary matrices $u_X$ and $u_\varphi$ that fulfill $u S_X = S_X u_X$ and $u S_\varphi = S_\varphi u_\varphi$.
    They can be constructed by either dropping rows and columns of $u$ corresponding to the null space, or taking the direct sum $u \oplus \unity$ with an identity matrix of a suitable dimension, depending on whether $\chi$ is larger or smaller than $\chi_X$ and $\chi_\varphi$.

    With the above technicalities out of the way, let us finish the proof.
    Choose $W = V_X u_X E V_\varphi\dg $.
    Then
    \begin{align}
        XW & = U_X S_X V_X\dg V_X u_X E V_\varphi\dg = U_X S_X u_X E V_\varphi\dg\\
        &= U_X u S_X E V_\varphi\dg = U_\varphi S_\varphi V_\varphi\dg = \varphi.
    \end{align}
    Thus we have found a $W$ such that $\varphi = XW$.
    To conclude the proof, we only need to observe that
    \begin{align}
        W W\dg & = V_X u_X E V_\varphi\dg V_\varphi E\dg u_X\dg V_X\dg\\
        & = V_X u_X E E\dg u_X\dg V_X\dg\\
        & = V_X u_X u\dg V_X\dg = V_X V_X\dg= \unity,
    \end{align}
    to see that $W$ is isometric.
\end{proof}

In the above, we have shown that if $\rho = XX\dg$, then any purification of $\rho$ can be written as $XW$ for some isometry $W$.
The only caveat here is the assumption that the ancilla space of the purification has a dimension $\chi_\varphi \ge \chi_X$.
This, however, can be easily circumvented by embedding any purification that uses a smaller ancilla, into a larger space.
Thus we conclude that Theorem~\ref{thrm:generic_purification} is a full characterization of all purifications of a density matrix of the form $\rho = XX\dg$.

\section{Solving the maximization}%
\label{app:one_isometry}
In this appendix, we show that for any matrix $M$,
\begin{align}
    \label{eq:one_isometry}
    \max_{W_1, W_2} \left| \Tr [ W_1 M W_2\dg ] \right| = \max_{W} \left| \Tr \left[ W M \right] \right| = \lVert M \rVert_{\text{tr}},
\end{align}
where $W_1$ and $W_2$ are isometric matrices in the sense that $W_1\dg W_1 = \unity$ and $W_2\dg W_2 = \unity$.
$W$ is also constrained to be isometric, and if we assume, with no loss of generality, that $M$ is of dimensions $\chi_1 \times \chi_2$ with $\chi_1 \le \chi_2$, then the isometricity condition on $W$ is $W\dg W = \unity$.

To get started, singular value decompose $M$ as $M = U S V\dg$.
Since $U$ and $V$ are unitary, we can always redefine $W_1 U \mapsto W_1$ and $W_2 V \mapsto W_2$ without affecting the isometricity of $W_1$ or $W_2$.
Thus,
\begin{align}
    \max_{W_1, W_2} \left| \Tr [ W_1 M W_2\dg ] \right| = \max_{W_1, W_2} \left| \Tr [ W_1 S W_2\dg ] \right|.
\end{align}
We can further rewrite
\begin{align}
    \label{eq:one_isometry_sum}
    \left| \Tr [W_1 S W_2\dg ] \right| = \left| \Tr [ S W_2\dg W_1 ] \right| = \left| \sum_{i=1}^{\rank M} \braket{W_{2,i}}{W_{1,i}} S_i \right|,
\end{align}
where $S_i$ are the singular values of $M$, and $\ket{W_{1,i}}$ and $\ket{W_{2,i}}$ are the $i^{\text{th}}$ columns of $W_1$ and $W_2$.
The isometricity of the $W$'s translates into
\begin{align}
    \braket{W_{1,i}}{W_{1,j}} = \delta_{ij} \quad \& \quad \braket{W_{2,i}}{W_{2,j}} = \delta_{ij}.
\end{align}
Given that $\ket{W_{1,i}}$ and $\ket{W_{2,i}}$ are normalized, and that $S_i$ are real and non-negative, it is clear that the best one can possibly hope to do in maximizing
\begin{align}
    \left| \sum_{i=1}^{\rank M} \braket{W_{2,i}}{W_{1,i}} S_i \right|,
\end{align}
is to have $\braket{W_{2,i}}{W_{1,i}} = 1$ for all $i$.
This can be achieved by choosing $\ket{W_{1,i}} = \ket{W_{2,i}}$ for the first $\rank M$ columns.
The remaining columns can be anything orthogonal to the first $\rank M$ ones, as they do not contribute, and similarly, what exactly $\ket{W_{1,i}}$ are chosen to be makes no difference.
With this choice we see that
\begin{align}
    \max_{W_1, W_2} \left| \Tr [ W_1 M W_2\dg ] \right| = \sum_{i=1}^{\rank M} S_i = \lVert M \rVert_{\text{tr}},
\end{align}
which proves the second part of Eq.~\eqref{eq:one_isometry}.

In what may seem like bizarrely over-complicated way of reexpressing the above result, we finally point out that with a logic very similar to the one above, one can easily show that
\begin{align}
    \max_{W} \left| \Tr [ W M ] \right| &= \max_{W} \left| \Tr [ W S  \right|\\
    &= \max_{W} \left| \sum_{i=1}^{\rank M} W_{ii} S_i \right|\\
    &= \sum_{i=1}^{\rank M} S_i = \lVert M \rVert_{\text{tr}}.
\end{align}
The usefulness of this first part of Eq.~\eqref{eq:one_isometry} can be found in the main text, where it is used to justify the definition of the quantity we call \emph{disjoint fidelity}.

\bibliography{uhlmann}

\end{document}